\documentclass[10pt,a4paper,twoside,pdf]{article}
\usepackage{geometry} % see geometry.pdf on how to lay out the page. There's lots.
\usepackage{graphicx}
\newcommand{\gae}
{\lower 2pt \hbox{$\, \buildrel {\scriptstyle >}\over {\scriptstyle \sim}\,$}}
\newcommand{\lae}
{\lower 2pt \hbox{$\, \buildrel {\scriptstyle <}\over {\scriptstyle \sim}\,$}}

\title{Newton's Old Bucket Experiment \& the Modern Liquid Telescope}
\author
{Aswin Sekhar$^{1,2*}$ \\
 $^1$Armagh Observatory, College Hill, Armagh BT61\ 9DG, United Kingdom\\
 $^2$Queen's University of Belfast, University Road, Belfast BT7 1NN, United Kingdom\\
 $^*$E-mail: asw@arm.ac.uk , asekhar01@qub.ac.uk \\ }

\date{} % delete this line to display the current date

\begin{document}

\maketitle{{\bf }

{ }
\section{Brief Introduction}

In nature there are various pairs of observed phenomena and observing scientific techniques which are elegantly coupled with each other. A very general and well known example is the fact that the metal we use to build telescopes were once built in stars by nuclear fusion. Hence in a fundamental sense, stars themselves have helped us indirectly in observing them in great detail. In this article I mention a bit more scientifically subtle and an even more interesting example of such a pair which raises interesting thoughts about an old experiment, its beauty and relevance in a very modern tool for astronomy. 

\section{Newton's Bucket Experiment}

This was one of the oldest simple experiments (after Galileo) which looked into interesting aspects of absolute and relative motion. Newton realised that any liquid rotating in a container would have a paraboloid shape and described about this experiment in 1689. Higher the rotational velocity, greater the depth of the paraboloid meniscus. The curvature of liquid surface seen while stirring coffee in a cup is exactly the same phenomenon. In our daily life, we usually experience this effect when the liquid rotates relative to the walls of a container.

Subsequently Newton imagined a scenario where the whole bucket and water together was spinning relative to the ground. In that case there is no relative motion between the liquid and walls of the bucket. For an observer on that spinning reference frame, both the water and bucket would remain stationary. Even then the observer would notice the parabolic shift in water's surface. Newton got inspired to find the origin and cause for such an effect mainly because the paraboloid shape is independent of the interaction between the container and liquid. 

Furthermore he imagined a case in which the universe had only a bucket of water. His intuitive calculations using classical mechanics predicted that the parabolic shift would be present even if there were no other perturbing masses in the universe with respect to which the bucket rotates (Page 78, Born \& Leibfried 1962; Narlikar 2011). In simple terms it meant that the paraboloid shape in any rotating liquid was a true signature of absolute rotation. Absolute motion with respect to space has been a subject of very thought provoking debates by great scientists like Newton, Descarte, Mach, Leibniz, Foucault, Neumann, Einstein and others (Chapter 7, Ciufolini \& Wheeler 1995). For the same experiment Mach was of the view that such a parabolic shift in a rotating fluid was due to the relative motion of the system with respect to other bodies in the universe and there is nothing called absolute rotation (Chapter 1, Barbour and Pfister 1995; Pfister and Braun 1985; Section 1, Raine 1975).

Centuries later Einstein verified both these hypothesis using his then newly invented general theory of relativity. Interestingly Einstein's calculations showed that Mach's argument was right in this particular context. He concluded that the parabolic shift is an inherent manifestation of rotation with respect to other celestial bodies and their interactions indeed contribute to this effect seen in the Newton's bucket experiment (Albert Einstein's letter to Ernst Mach dated 25 June 1913; Chapter 21, Section 12, Misner et al. 1973). In simple words, it meant that the gravitational interaction of stars plays a key role in the phenomenon of paraboloid shape in any rotating fluid system. 

The scientific as well as philosophical aspects of this phenomena has been a subject of widespread interest and study (Chapter 1, DiSalle 2006) by many pioneers in classical mechanics and general relativity.

\section{Modern Liquid Telescope}

The idea of using a liquid as a primary mirror dates back to Ernesto Capocci's letter (Gibson 1991) to the Royal Academy of Belgium in 1850. It never became a reality then because of the lack of efficient electric motors to rotate the system and thereby generate the required parabolic shift (discussed above) in a liquid to make it act as a concave mirror of the right focal length. But there was widespread interest (Borra et al. 1991, Hickson 2002) among many group of scientists ever since then. American physicist Robert W. Wood built one of the first complete liquid mirror telescopes (Gibson 1991) in 1909. With his 0.51 m mirror, he was able to resolve the e Lyrae quadruple star system, which has component stars separated by as small as 2.3 arc seconds.

The Large Zenith Telescope (LZT) completed in 2003 has proved to be remarkably successful and intriguing because of its novel mechanism and high efficiency. It uses the reflective liquid Mercury to focus the starlight. The 6 metre dish with 30 litres of Mercury spins at about 7 times per minute which in turn gives the right curvature to the fluid to act as a perfect concave mirror. The overall cost of this project has been 10 times lower than that of the telescopes with conventional solid mirrors. The success of this project has inspired many international institutes to plan projects of similar magnitudes and higher. For example, University of British Columbia is planning a 10 metre liquid mirror telescope in near future in close collaboration with America and Australia using the same old technique outlined in Newton's bucket experiment. 

\section{The Beautiful Convergence}

Section 2 and 3 brings us back to the original comparison similar to the case mentioned in the introduction where stardust in turn aids to observe stars. In this specific case, the sheer concept of observing the celestial bodies using the parabolic shift of a rotating fluid which in turn is believed to be caused by the gravitational effects from these celestial bodies itself gives an unusual scientific and philosophical appeal to the whole picture. Hence the sudden advancements and flourishing of many liquid telescope projects give a new twist to the entire scene. It simply makes the Newtonian, Machian and Einsteinian view of absolute motion and rotating reference frames even more relevant and thought provoking in the present times.

\section{References}

\ \ \ \ 1. Barbour J. and Pfister H. 1995. Mach's Principle - From Newton's Bucket to Quantum Gravity. Birkhauser. Boston.\

2. Born M. and Leibfried G. 1962. Einstein's Theory Of Relativity. Dover Publications. New York.\

3. Borra E. F. 1982. The Liquid-Mirror Telescope as a Viable Astronomical Tool. Journal of Royal Astronomical Society of Canada 76, 245.\

4. Borra E. F. 1991. The Case for a Liquid Mirror in a Lunar based Telescope. The Astrophysical Journal 373, 317.\

5. Ciufolini I. and Wheeler J. A. 1995. Gravitation and Inertia. Princeton University Press. \

6. Disalle R. 2002. The Cambridge Companion to Newton. Cambridge University Press.\

7. Gibson B. K. 1991. "Liquid Mirror Telescopes: History". Journal of Royal Astronomical Society of Canada, 85, 158.\

8. Hickson P. 2002. Wide-Field Tracking with Zenith-Pointing Telescopes. Monthly Notices of Royal Astronomical Society 330, 540.\

9. Misner C. W., Thorne, K. S. and Wheeler J.  A. 1973. Gravitation. Freeman. New York.\

10. Narlikar J. V. 2011. Resonance (Journal of Indian Academy of Sciences) 16, 310.\

11. Pfister H. and Braun K. H. 1985. Induction of Correct Centrifugal Force in a Rotating Mass Shell. Classical and Quantum Gravity 2, 909.\

12. Raine D. J. 1975. Mach's Principle in General Relativity. Monthly Notices of Royal Astronomical Society 171, 507.\

\end{document}